\documentclass[10pt]{article}
\usepackage{graphicx}
\usepackage{amsmath}
\usepackage{amssymb}
\usepackage{caption2}
\begin{document}
\bibliographystyle{prsty}
\begin{center}
{\large {\bf \sc{  The radiative decays  $B_c^{*\pm} \to B_c^{\pm} \gamma $ with QCD sum rules }}} \\[2mm]
Zhi-Gang Wang \footnote{E-mail,zgwang@aliyun.com.  }     \\
 Department of Physics, North China Electric Power University,
Baoding 071003, P. R. China
\end{center}

\begin{abstract}
In this article, we calculate the $B_c^* \to B_c$ electromagnetic form-factor with the three-point QCD sum rules,  then  study
 the radiative decays $ B_c^{*\pm} \to B_c^{\pm} \gamma$.  Experimentally,   we can study the radiative transitions    using the decay cascades  $B_c^{*\pm}\to B_c^{\pm} \gamma\to J/\psi \ell^{\pm}\bar{\nu}_{\ell} \gamma\to \mu^+ \mu^- \ell^{\pm}\bar{\nu}_{\ell} \gamma$   in the future at the LHCb.
\end{abstract}



\section{Introduction}
The ground state bottom-charm  mesons,  which lie below the $BD$, $BD^*$, $B^*D$, $B^*D^*$ thresholds, cannot annihilate into gluons due to their flavor composing, and decay weakly through  $\bar{b} \to \bar{c}W^+$, $c \to s W^+$, $c\bar{b}\to W^+$ or decay radiatively through $b \to b \gamma$, $c \to c \gamma$
 at the quark level.  The pseudoscalar mesons $B_c^{\pm}$  decay weakly and  have measurable lifetime, while the radiative transitions $B_c^{*\pm} \to B_c^{\pm}\gamma$ saturate the widths of the vector mesons $B_c^{*\pm}$.
 Experimentally, the semileptonic decays $B_c^{\pm} \to J/\psi \ell^{\pm}\bar{\nu}_{\ell}$, $B_c^{+} \to J/\psi e^{+}\bar{\nu}_{e} $ were   used to measure the $B_c$ lifetime and the hadronic decays $B_c^{\pm} \to J/\psi \pi^{\pm}$ were used to measure the $B_c$ mass \cite{BcExp}.
  The $B_c^{*\pm}$ mesons have not been observed yet, but they are expected to be observed  at the large hadron collider (LHC) through the radiative transitions.
In the article, we calculate the $B_c^* \to B_c$ electromagnetic form-factor with the three-point QCD sum rules, and study the radiative decays $B_c^{*\pm} \to B_c^{\pm}\gamma$.

The QCD sum rules is a powerful nonperturbative approach   in
 studying   the heavy quarkonium states, and has given many successful descriptions of the masses, decay constants, form-factors, strong coupling constants  \cite{SVZ79,Reinders85,QCDSR-review}.
 The weak form-factors  $B_c  \to    J/\psi$, $\eta_c$, $\chi_{c0}$, $\chi_{c1}$, $h_c$, $B$, $B_s$, $D$, $D_s$ , $B^*$, $B^*_s$, $D^*$, $D^*_s$, $D_{s1}$, etc, have been studied extensively with the three-point QCD sum rules \cite{Bagan1994,Colangelo1993,Kiselev2000,Coulomb-BC,QCDSR-BC}, and the corresponding semileptonic decay widths have also been studied. In previous work, we calculate the $B_c^* \to \eta_c$ form-factors with the three-point QCD sum rules, and study   the semileptonic decays  $B_c^* \to \eta_c \ell \bar{\nu}_{\ell}$ \cite{Wang1209}. The tiny decay widths are consistent with the expectation  that the radiative transitions $B_c^{*\pm} \to B_c^{\pm}\gamma$ have the dominant branching fractions. In the past years, the radiative transitions $B_c^{*\pm} \to B_c^{\pm}\gamma$ have been studied by the (non-) relativistic   potential models  \cite{GKLT,EQ,Fulc,EFG,Godfrey2004}. It is interesting to make prediction based on the  nonperturbative method of QCD.

The article is arranged as follows:  we study the $B_c^* \to B_c  $ electromagnetic form-factor  using
  the three-point QCD sum rules in Sect.2; in Sect.3, we present the numerical results and discussions; and Sect.4 is reserved for our
conclusions.

\section{ The $B_c^* \to B_c$ electromagnetic form-factor with QCD sum rules}
We study the $B_c^* \to B_c$ electromagnetic form-factor with  the three-point correlation function $\Pi_{\mu\nu}(p_1,p_2)$,
\begin{eqnarray}
\Pi_{\mu\nu}(p_1,p_2)&=&i^2\int d^4 x d^4 y e^{ip_2\cdot x-ip_1\cdot y} \langle 0|T\{J_5(x) j_\mu (0) J_\nu^{\dagger}(y)\} |0\rangle \, ,
\end{eqnarray}
where
\begin{eqnarray}
j_\mu(0)&=&e_b\bar{b}(0)\gamma_\mu b(0)+e_c\bar{c}(0)\gamma_\mu c(0)\, , \nonumber\\
J_5(x)&=&\bar{c}(x)i\gamma_5b(x)\, , \nonumber\\
J_\nu(y)&=&\bar{c}(y)\gamma_\nu b(y)\, ,
\end{eqnarray}
the $j_\mu(0)$ is the electromagnetic current, the electric charges $e_b=-\frac{1}{3}$ and $e_c=\frac{2}{3}$, the currents  $J_5(x)$ and  $J_\nu(y)$ interpolate the pseudoscalar  and vector  $B_c$ mesons, respectively.

We can insert  a complete set of intermediate hadronic states with
the same quantum numbers as the current operators $J_5(x)$ and $J^{\dagger}_\nu(y)$ into the
correlation function $\Pi_{\mu\nu}(p_1,p_2)$  to obtain the hadronic representation
\cite{SVZ79,Reinders85}. After isolating the ground state
contributions come from the heavy mesons $B_c^*$ and $B_c$ , we get the following result,
\begin{eqnarray}
\Pi_{\mu\nu}(p_1,p_2)&=&\frac{\langle 0|J_5(0) |B_c(p_2)\rangle \langle B_c(p_2)|j_\mu(0) |B_c^*(p_1)\rangle \langle B_c^*(p_1)  |J^{\dagger}_\nu(0) |0   \rangle}{(M_{B_c}^2-p_2^2)(M_{B_c^*}^2-p_1^2)} +\cdots   \, ,\nonumber\\
&=&-\frac{f_{B_c}M_{B_c}^2f_{B_c^*}M_{B_c^*}V(q^2)}{(m_b+m_c)(M_{B_c^*}+M_{B_c})(M_{B_c^*}^2-p_1^2)(M_{B_c}^2-p_2^2)}  \epsilon_{\mu\nu\alpha\beta}p_1^\alpha p_2^\beta +\cdots \, ,
\end{eqnarray}
where we have used the following definitions for the $B_c^* \to B_c$ electromagnetic form-factor and weak decay constants of the vector meson $B_c^*$ and pseudoscalar meson $B_c$,
\begin{eqnarray}
\langle B_c(p_2)|j_\mu(0) |B_c^*(p_1)\rangle&=& \epsilon_{\mu\nu\alpha\beta}\varepsilon^{\nu}p_1^\alpha p_2^\beta \frac{V(q^2)}{M_{B_c^*}+M_{B_c}}\, ,\\
\langle0|J_\mu(0) |B_c^*(p_1)\rangle&=&f_{B_c^*}M_{B_c^*}\varepsilon_\mu \, ,\nonumber\\
\langle0|J_5(0) |B_c(p_2)\rangle&=&\frac{f_{B_c}M_{B_c}^2}{m_b+m_c} \, ,
\end{eqnarray}
$q_\mu=(p_1-p_2)_\mu$, the $\varepsilon_\mu$ is the polarization  vector of the $B_c^*$ meson.

Now, we briefly outline  the operator product expansion for the correlation function $\Pi_{\mu\nu}(p_1,p_2)$.  We contract the quark fields in the correlation function
$\Pi_{\mu\nu}(p_1,p_2)$ with Wick theorem firstly,
\begin{eqnarray}
\Pi_{\mu\nu}(p_1,p_2)&=&\int d^4 x d^4 y e^{ip_2\cdot x-ip_1\cdot y} \left\{e_b{\rm Tr} \left[i\gamma_5B^{mn}(x)\gamma_{\mu}B^{nk}(-y)\gamma_\nu C^{km}(y-x) \right] \right. \nonumber\\
&&\left.+ e_c{\rm Tr} \left[i\gamma_5B^{mn}(x-y)\gamma_{\nu}C^{nk}(y)\gamma_\mu C^{km}(-x) \right] \right\} \, ,
\end{eqnarray}
replace the $c$ and $b$ quark propagators $C^{ij}(x) $ and $B^{ij}(x)$ with the corresponding full propagators $S_{ij}(x)$,
\begin{eqnarray}
S_{ij}(x)&=&\frac{i}{(2\pi)^4}\int d^4k e^{-ik \cdot x} \left\{
\frac{\delta_{ij}}{\!\not\!{k}-m_Q}
-\frac{g_sG^n_{\alpha\beta}t^n_{ij}}{4}\frac{\sigma^{\alpha\beta}(\!\not\!{k}+m_Q)+(\!\not\!{k}+m_Q)
\sigma^{\alpha\beta}}{(k^2-m_Q^2)^2}+\frac{\delta_{ij}\langle g^2_sGG\rangle }{12}\right.\nonumber\\
&&\left. \frac{m_Qk^2+m_Q^2\!\not\!{k}}{(k^2-m_Q^2)^4}+\cdots\right\} \, ,
\end{eqnarray}
 where $Q=c,b$, $t^n=\frac{\lambda^n}{2}$, the $\lambda^n$ are the Gell-Mann matrixes, the $i$, $j$ are color indexes, and the $\langle g^2_sGG\rangle$
is the gluon condensate \cite{Reinders85},
then carry out  the  integrals. In this article, we take into account the leading-order perturbative contribution $\Pi_{\mu\nu}^{0}(p_1,p_2)$ and  gluon condensate contribution $\Pi_{\mu\nu}^{GG}(p_1,p_2)$ in the operator product expansion.

The leading-order perturbative contribution $\Pi_{\mu\nu}^{0}(p_1,p_2)$ can be written as
\begin{eqnarray}
\Pi_{\mu\nu}^{0}(p_1,p_2)&=&\frac{3e_b}{(2\pi)^4}\int d^4k \frac{ {\rm Tr}\left\{ \gamma_5\left[ \!\not\!{k}+ \!\not\!{p}_2+m_b\right]\gamma_\mu \left[ \!\not\!{k}+ \!\not\!{p}_1+m_b\right]\gamma_\nu\left[ \!\not\!{k} +m_c\right]\right\}}{\left[(k+p_2)^2-m_b^2\right]\left[(k+p_1)^2-m_b^2\right]\left[k^2-m_c^2\right]}\, ,\nonumber\\
&&+\frac{3e_c}{(2\pi)^4}\int d^4k \frac{ {\rm Tr}\left\{ \gamma_5\left[ \!\not\!{k}+  m_b\right]\gamma_\nu \left[ \!\not\!{k}- \!\not\!{p}_1+m_c\right]\gamma_\mu\left[ \!\not\!{k} -\!\not\!{p}_2+m_c\right]\right\}}{\left[k^2-m_b^2\right]\left[(k-p_1)^2-m_c^2\right]\left[(k-p_2)^2-m_c^2\right]}\, ,\nonumber\\
&=&\int ds_1 ds_2 \frac{\rho_{\mu\nu}(s_1,s_2,q^2)}{(s_1-p_1^2)(s_2-p_2^2)} \, .
\end{eqnarray}
We put all the quark lines on mass-shell using the Cutkosky's rule,
and  obtain the leading-order perturbative spectral density  $\rho_{\mu\nu} (s_1,s_2,q^2)$,
\begin{eqnarray}
\rho_{\mu\nu}(s_1,s_2,q^2)  &=&-\frac{3ie_b}{(2\pi)^3} \int d^4k \delta\left[(k+p_2)^2-m_b^2\right]\delta\left[(k+p_1)^2-m_b^2\right]\delta\left[k^2-m_c^2\right]\nonumber\\
&& {\rm Tr}\left\{ \gamma_5\left[ \!\not\!{k}+ \!\not\!{p}_2+m_b\right]\gamma_\mu\left[ \!\not\!{k}+ \!\not\!{p}_1+m_b\right]\gamma_\nu\left[ \!\not\!{k} +m_c\right]\right\} \nonumber\\
&&-\frac{3ie_c}{(2\pi)^3} \int d^4k \delta\left[(k-p_2)^2-m_c^2\right]\delta\left[(k-p_1)^2-m_c^2\right]\delta\left[k^2-m_b^2\right]\nonumber\\
&& {\rm Tr}\left\{ \gamma_5\left[ \!\not\!{k} +m_b\right]\gamma_\nu\left[ \!\not\!{k}-\!\not\!{p}_1+m_c\right]\gamma_\mu\left[ \!\not\!{k} -\!\not\!{p}_2 +m_c\right]\right\}\nonumber\\
&=& -\frac{3e_b\epsilon_{\mu\nu\alpha\beta}p_1^\alpha p_2^\beta}{ 4\pi^2\sqrt{\lambda(s_1,s_2,q^2)}}\left\{m_b +\frac{(m_b-m_c)(s_1+s_2-q^2+2m_b^2-2m_c^2)q^2}{\lambda(s_1,s_2,q^2)}\right\} \nonumber\\
&& -\frac{3e_c\epsilon_{\mu\nu\alpha\beta}p_1^\alpha p_2^\beta}{ 4\pi^2\sqrt{\lambda(s_1,s_2,q^2)}}\left\{m_c +\frac{(m_c-m_b)(s_1+s_2-q^2+2m_c^2-2m_b^2)q^2}{\lambda(s_1,s_2,q^2)}\right\}  \, ,
\end{eqnarray}
$\lambda(a,b,c)=a^2+b^2+c^2-2ab-2bc-2ca$, where we have  used the formulae presented in Refs.\cite{Wang1209,Gongshi-Ioffe}
to carry out the integrals.

 We calculate the gluon condensate contribution directly and obtain the following expression,
\begin{eqnarray}
\Pi_{\mu\nu}^{GG}(p_1,p_2)&=&\frac{ie_b\epsilon_{\mu\nu\alpha\beta}}{4\pi^2} \langle\frac{\alpha_sGG}{\pi}\rangle \left\{ -m_b^2m_c \left(\overline{I}_{141}+\overline{I}_{411}\right)p_1^{\alpha} p_2^\beta \right.\nonumber\\
&&+m_b^2(m_c-m_b)\left(\overline{I}^\alpha_{141}+\overline{I}^\alpha_{411}\right)q^\beta-m_c^3\overline{I}_{114}p_1^\alpha p_2^\beta+m_c^2(m_c-m_b)\overline{I}^\alpha_{114}q^\beta\nonumber\\
&&\left.-m_b\overline{I}^\alpha_{131}p_1^\beta+m_b\overline{I}^\alpha_{311}p_2^\beta
-m_c\overline{I}_{113}p_1^\alpha p_2^\beta+m_c\overline{I}_{113}^\alpha q^\beta \right\}\nonumber\\
&&+\frac{ie_b\epsilon_{\mu\nu\alpha\beta}}{24\pi^2} \langle\frac{\alpha_sGG}{\pi}\rangle \left\{ (m_b-m_c)\left(\overline{I}_{221}^\alpha+\overline{I}_{212}^\alpha\right) q^\beta +m_c\left(\overline{I}_{221}+\overline{I}_{212}\right)p_1^\alpha p_2^\beta\right. \nonumber\\
&&\left. -2m_b\overline{I}_{122}^\alpha p_2^\beta-3(m_b-m_c)\overline{I}_{122}^\alpha q^\beta-3m_c \overline{I}_{122} p_1^\alpha p_2^\beta \right\} \nonumber\\
&&+\frac{ie_c\epsilon_{\mu\nu\alpha\beta}}{4\pi^2} \langle\frac{\alpha_sGG}{\pi}\rangle \left\{ -m_c^2m_b \left(\overline{J}_{141}+\overline{J}_{411}\right)p_1^{\alpha} p_2^\beta \right.\nonumber\\
&&+m_c^2(m_b-m_c)\left(\overline{J}^\alpha_{141}+\overline{J}^\alpha_{411}\right)q^\beta-m_b^3\overline{J}_{114}p_1^\alpha p_2^\beta+m_b^2(m_b-m_c)\overline{J}^\alpha_{114}q^\beta\nonumber\\
&&\left.-m_c\overline{J}^\alpha_{131}p_1^\beta+m_c\overline{J}^\alpha_{311}p_2^\beta
-m_b\overline{J}_{113}p_1^\alpha p_2^\beta+m_b\overline{J}_{113}^\alpha q^\beta \right\}\nonumber\\
&&+\frac{ie_c\epsilon_{\mu\nu\alpha\beta}}{24\pi^2} \langle\frac{\alpha_sGG}{\pi}\rangle \left\{ (m_c-m_b)\left(\overline{J}_{221}^\alpha+\overline{J}_{212}^\alpha\right) q^\beta +m_b\left(\overline{J}_{221}+\overline{J}_{212}\right)p_1^\alpha p_2^\beta\right. \nonumber\\
&&\left. -2m_c\overline{J}_{122}^\alpha p_2^\beta-3(m_c-m_b)\overline{J}_{122}^\alpha q^\beta-3m_b \overline{J}_{122} p_1^\alpha p_2^\beta \right\}\, ,
\end{eqnarray}
where
\begin{eqnarray}
\overline{I}_{ijn}&=& \int d^4k \frac{1}{\left[(k+p_1)^2-m_b^2\right]^i \left[(k+p_2)^2-m_b^2\right]^j \left[ k^2-m_c^2\right]^n} \, , \nonumber\\
\overline{I}_{ijn}^\alpha&=& \int d^4k \frac{k^\alpha}{\left[(k+p_1)^2-m_b^2\right]^i \left[(k+p_2)^2-m_b^2\right]^j \left[ k^2-m_c^2\right]^n} \, , \nonumber\\
\overline{J}_{ijn}&=& \int d^4k \frac{1}{\left[(k+p_1)^2-m_c^2\right]^i \left[(k+p_2)^2-m_c^2\right]^j \left[ k^2-m_b^2\right]^n} \, , \nonumber\\
\overline{J}_{ijn}^\alpha&=& \int d^4k \frac{k^\alpha}{\left[(k+p_1)^2-m_c^2\right]^i \left[(k+p_2)^2-m_c^2\right]^j \left[ k^2-m_b^2\right]^n} \, .
\end{eqnarray}

 We take  quark-hadron duality  below the threshold
$s^0_{1}$ and $s_2^0$ in the channels $B_c^*$ and $B_c$, respectively,
 perform double Borel transform  with respect to the variables
$P_1^2=-p_1^2$ and $P_2^2=-p_2^2$, respectively,
and  obtain the QCD  sum rule  for the electromagnetic  form-factor $V(q^2)$,
 \begin{eqnarray}
 V(q^2)&=&\frac{(M_{B_c^*}+M_{B_c})(m_b+m_c)}{f_{B_c^*}f_{B_c}M_{B_c^*}M_{B_c}^2}\exp\left( \frac{M_{B_c^*}^2}{M_1^2}+\frac{M_{B_c}^2}{M_2^2}\right)\nonumber\\
 &&\left\{ \frac{3e_b}{ 4\pi^2}\int_{(m_b+m_c)^2}^{s_1^0} ds_1 \int_{(m_b+m_c)^2}^{s_2^0} ds_2 \frac{\mathcal{C}}{\sqrt{\lambda(s_1,s_2,q^2)}}\exp\left( -\frac{s_1}{M_1^2}-\frac{s_2}{M_2^2}\right) \right.\nonumber\\
&&\left[m_b +\frac{(m_b-m_c)(s_1+s_2-q^2+2m_b^2-2m_c^2)q^2}{\lambda(s_1,s_2,q^2)}\right]|_{|b(s_1,s_2,q^2)|\leq 1} \nonumber\\
&& +\frac{3e_c }{  4\pi^2}\int_{(m_b+m_c)^2}^{s_1^0} ds_1 \int_{(m_b+m_c)^2}^{s_2^0} ds_2 \frac{\mathcal{C}}{\sqrt{\lambda(s_1,s_2,q^2)}}\exp\left( -\frac{s_1}{M_1^2}-\frac{s_2}{M_2^2}\right)\nonumber\\
&&\left.\left[m_c +\frac{(m_c-m_b)(s_1+s_2-q^2+2m_c^2-2m_b^2)q^2}{\lambda(s_1,s_2,q^2)}\right]|_{|c(s_1,s_2,q^2)|\leq 1}  \right\}\ \nonumber\\
&&-\frac{(M_{B_c^*}+M_{B_c})(m_b+m_c)M_1^2M_2^2}{f_{B_c^*}f_{B_c}M_{B_c^*}M_{B_c}^2}\langle\frac{\alpha_sGG}{\pi}\rangle \exp\left( \frac{M_{B_c^*}^2}{M_1^2}+\frac{M_{B_c}^2}{M_2^2}\right) \nonumber\\
&&\left\{ \frac{e_b m_b^2}{4\pi^2}\left[m_c\left(I_0^{141}+I_0^{411}\right)  +(m_c-m_b)\left(I_{10}^{141}+I_{01}^{141}+I_{10}^{411}+I_{01}^{411} \right)\right] \right. \nonumber\\
&&+ \frac{e_b m_c^2}{4\pi^2}\left[m_c I_0^{114}   +(m_c-m_b)\left(I_{10}^{114}+I_{01}^{114} \right)\right]-\frac{e_bm_b}{4\pi^2}\left(I_{01}^{131}+I_{10}^{311} \right)\nonumber\\
&&+\frac{e_b m_c}{4\pi^2}\left(I_{0}^{113}+I_{10}^{113}+I_{01}^{113} \right)+\frac{e_b(m_b-m_c)}{24\pi^2}\left( I_{10}^{221}+I_{01}^{221}+I_{10}^{212}+I_{01}^{212}\right) \nonumber\\
&&+\frac{e_bm_b}{12\pi^2}I_{10}^{122}-\frac{e_b(m_b-m_c)}{8\pi^2}\left(I_{10}^{122}+I_{01}^{122} \right)-\frac{e_b m_c}{24\pi^2}\left( I_0^{221}+I_{0}^{212}-3I_0^{122}\right) \nonumber\\
&&+\frac{e_c m_c^2}{4\pi^2}\left[m_b\left(J_0^{141}+J_0^{411}\right)  +(m_b-m_c)\left(J_{10}^{141}+J_{01}^{141}+J_{10}^{411}+J_{01}^{411} \right)\right] \nonumber\\
&&+ \frac{e_c m_b^2}{4\pi^2}\left[m_b J_0^{114}   +(m_b-m_c)\left(J_{10}^{114}+J_{01}^{114} \right)\right]-\frac{e_c m_c}{4\pi^2}\left(J_{01}^{131}+J_{10}^{311} \right)\nonumber\\
&&+\frac{e_c m_b}{4\pi^2}\left(J_{0}^{113}+J_{10}^{113}+J_{01}^{113} \right)+\frac{e_c(m_c-m_b)}{24\pi^2}\left( J_{10}^{221}+J_{01}^{221}+J_{10}^{212}+J_{01}^{212}\right) \nonumber\\
&&\left.+\frac{e_c m_c}{12\pi^2}J_{10}^{122}-\frac{e_c(m_c-m_b)}{8\pi^2}\left(J_{10}^{122}+J_{01}^{122} \right)-\frac{e_c m_b}{24\pi^2}\left( J_0^{221}+J_{0}^{212}-3J_0^{122}\right) \right\}  \, ,\\
&=&e_b V_1(q^2)+e_c V_2(q^2)\, ,
\end{eqnarray}
where
\begin{eqnarray}
b(c_1,c_2,q^2) &=& \frac{2s_1(s_2+m_c^2-m_b^2)-(s_1+s_2-q^2)(s_1+m_c^2-m_b^2)}{\sqrt{\lambda(s_1,s_2,q^2)\lambda(s_1,m_c^2,m_b^2)}}  \, , \nonumber\\
c(c_1,c_2,q^2) &=& \frac{2s_1(s_2+m_b^2-m_c^2)-(s_1+s_2-q^2)(s_1+m_b^2-m_c^2)}{\sqrt{\lambda(s_1,s_2,q^2)\lambda(s_1,m_c^2,m_b^2)}}  \, , \nonumber\\
{\mathcal{C}}&=&\sqrt{\frac{4\pi\alpha_s^\mathcal{C}}{3v_1} \left[1-\exp\left(-\frac{4\pi\alpha_s^\mathcal{C}}{3v_1}\right)\right]^{-1}}\sqrt{\frac{4\pi\alpha_s^\mathcal{C}}{3v_2} \left[1-\exp\left(-\frac{4\pi\alpha_s^\mathcal{C}}{3v_2}\right)\right]^{-1}}\, ,\nonumber\\
v_1&=&\sqrt{1-\frac{4m_bm_c}{s_1-(m_b-m_c)^2}}\, ,\nonumber\\
v_2&=&\sqrt{1-\frac{4m_bm_c}{s_2-(m_b-m_c)^2}}\, ,\nonumber\\
\end{eqnarray}
the explicit expressions of the $I_0^{ijn}$, $I_{10}^{ijn}$, $I_{01}^{ijn}$, $J_0^{ijn}$, $J_{10}^{ijn}$, $J_{01}^{ijn}$ are presented in the appendix.
For the heavy quarkonium states $B_c^*$ and $B_c$, the relative velocities of quark movement are small, we should account for the Coulomb-like $\frac{\alpha_s^\mathcal{C}}{v}$  corrections.  After taking into account  all the Coulomb-like contributions shown in Fig.1, we obtain the coefficient $\mathcal{C}$ to dress the   quark-meson  vertex \cite{Kiselev2000,Coulomb-BC}.

At the recoil momentum close to zero, the heavy quark velocities are small below the thresholds $s_1^0$ and $s_2^0$,
the ladder Feynman diagrams shown in Fig.1 are calculated  in the nonrelativistic approximation, and result in the coefficient  $\mathcal{C}$ to dress the
 quark-meson  vertex.  In our previous work on the two-point QCD sum rules for the $B_c^*$ mesons \cite{Wang-BC}, we observed that the  perturbative ${\mathcal{O}}(\alpha_s)$ corrections to the leading-order spectral density $\rho_{0}(s)$ can be  approximated  by $\rho_0(s)\frac{2\pi\alpha_s^{\mathcal{C}}}{3v}$ with
the assumption  ${\alpha_s^\mathcal{C}}=\alpha_s(\mu)$, and
  accounted for all the   Coulomb-like contributions (or all the perturbative corrections approximately) by multiplying the $\rho_{0}(s)$ with the coefficient $\mathcal{C}$,
\begin{eqnarray}
{\mathcal{C}}&=&\frac{4\pi\alpha_s^\mathcal{C}}{3v} \frac{1}{1-\exp\left(-\frac{4\pi\alpha_s^\mathcal{C}}{3v}\right)}=1+\frac{2\pi\alpha_s^\mathcal{C}}{3v}+\frac{1}{12}\left(\frac{4\pi\alpha_s^\mathcal{C}}{3v}\right)^2
-\frac{1}{720}\left(\frac{4\pi\alpha_s^\mathcal{C}}{3v}\right)^4+\cdots\,   .
\end{eqnarray}
In the case of the three-point QCD sum rules, the perturbative ${\mathcal{O}}(\alpha_s)$ corrections to the leading order spectral densities are available only for the electromagnetic form-factors of the $\pi$ and $\rho$ mesons \cite{pion-rho}, we expect to approximate the perturbative ${\mathcal{O}}(\alpha_s)$ corrections by multiplying the leading order spectral densities with $\frac{\pi\alpha_s^{\mathcal{C}}}{3v_1}+\frac{\pi\alpha_s^{\mathcal{C}}}{3v_2}$, and take into account all the Coulomb-like interactions (or all the perturbative corrections approximately)  by   multiplying the leading order spectral densities with the coefficient $\mathcal{C}$ \cite{Kiselev2000,Coulomb-BC}. Direct but formidable calculations of the perturbative  corrections are still  needed to validate or invalidate the present approximation.
 In the region of physical resonances,  the most essential
effect comes from the normalization factor $\mathcal{C}$.  In the case of the two-point sum rules,  the normalization factor $\mathcal{C}$  leads to a double-triple multiplication of the tree-level value of the spectral densities numerically \cite{PRT1978}.
The  coefficient $\mathcal{C}$ survives beyond the zero recoil limit,
 or at least serve as  upper bounds on the form-factors  in the QCD sum rules  \cite{Kiselev2000,Coulomb-BC}.
  In this article, we take the approximation ${\alpha_s^\mathcal{C}}=\alpha_s(\mu)$ in numerical  calculation as in our previous work \cite{Wang-BC}.

\begin{figure}
 \centering
 \includegraphics[totalheight=3.5cm,width=6cm]{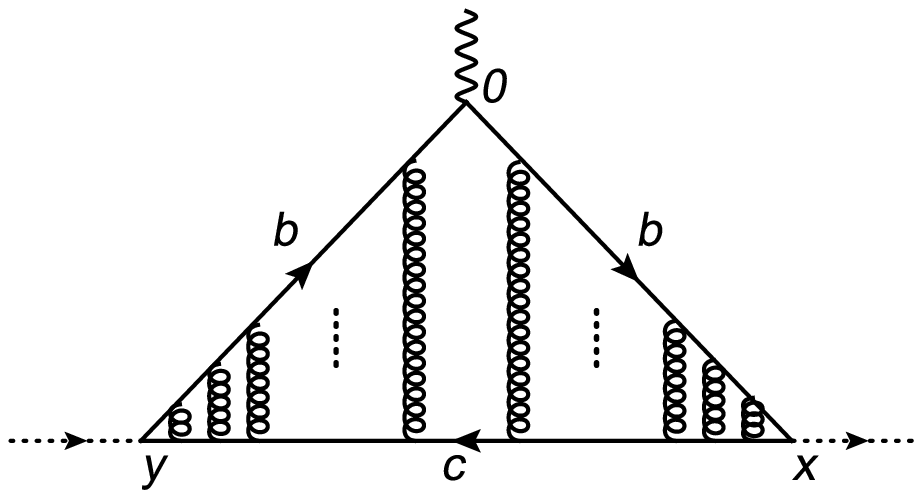}
  \includegraphics[totalheight=3.5cm,width=6cm]{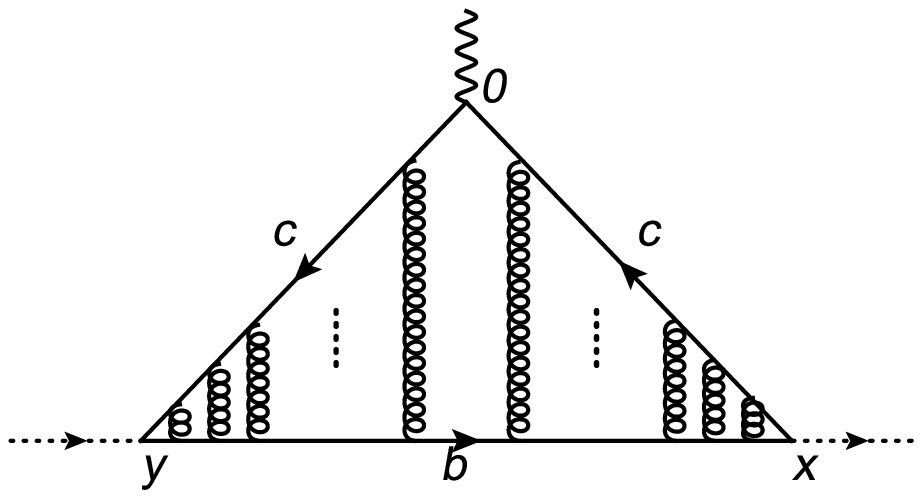}
    \caption{The ladder Feynman diagrams for the Coulomb-like interactions. }
\end{figure}

In the physical region $q^2=0$, the constraints $|b(c_1,c_2,0)| \leq 1  $ and $|c(c_1,c_2,0)| \leq 1  $ lead to the inequations,
\begin{eqnarray}
 -1\leq \frac{s_1+(m_b^2-m_c^2)}{\sqrt{\lambda(s_1,m_1^2,m_2^2)}} \leq 1 \, ,\nonumber\\
 -1\leq \frac{s_1+(m_c^2-m_b^2)}{\sqrt{\lambda(s_1,m_1^2,m_2^2)}} \leq 1 \, ,
\end{eqnarray}
those constraints cannot be satisfied.  In this article, we calculate the electromagnetic form-factors $V_1(q^2)$ and $V_2(q^2)$
at the space-like region $Q^2=-q^2=(1.0-5.4)\,\rm{GeV}^2$, then fit the electromagnetic form-factors  with suitable analytical functions, and obtain the value $V(0)$ by analytically  continuing  the variable $q^2$ to the physical region.

\section{Numerical results and discussions}

  The pseudoscalar  mesons $B_c$ have been studied by  the full QCD sum rules \cite{Bagan1994,Colangelo1993,Chabab1994,Narison1988} and the potential approach
combined with the QCD sum rules \cite{GKLT,Kiselev1993}, while the vector mesons $B_c^*$ have been studied by the full QCD sum rules \cite{Colangelo1993,Wang-BC,Narison1988}.
The predictions for the masses and decay constants are
$f_{B_c}=(0.383\pm 0.027)\,\rm{GeV}$ \cite{Bagan1994};
$f_{B_c^*}\approx f_{B_c}=(360\pm60)\,\rm{MeV}$,  $M_{B_c}\approx 6.35\,\rm{GeV}$ \cite{Colangelo1993};
$f_{B_c}=(460 \pm 60)\,\rm{MeV}$ \cite{GKLT}; $f_{B_c^*}=(0.384 \pm 0.032)\,\rm{GeV}$, $M_{B_c^*}=
(6.337 \pm 0.052)\,\rm{GeV}$ \cite{Wang-BC};
$f_{B_c}=(300\pm65)\, \rm{MeV}$, $M_{B_c}=(6.25\pm0.20)\,\rm{GeV}$ \cite{Chabab1994};
  $f_{B_c}=(566\pm28)\,\rm{MeV}$, $f_{B_c^*}\approx (0.346\pm0.025)\,\rm{GeV}$, $M_{B_c} \approx(6.969\pm0.18)\,\rm{GeV}$,
$M_{B_c^*} \approx(6.855\pm0.18)\,\rm{GeV}$ \cite{Narison1988}.   The predictions for the mass $M_{B_c}$ are consistent with (or much larger than) the average value $M_{B_c}=(6.277 \pm 0.006)\,\rm{GeV}$ listed in the  Review of Particle Physics \cite{PDG}, while the predictions for the decay constant $f_{B_c}$ vary in large ranges.

The values of the decay constants from other theoretical calculations  also vary in large ranges,
$f_{B^*_c}=f_{B_c}=500\,\rm{MeV}$, $512\,\rm{MeV}$, $479\,\rm{MeV}$ and $687\,\rm{MeV}$ from the Buchmuller-Tye potential,   power-law potential,  logarithmic potential and Cornell potential, respectively \cite{EQ};
 $f_{B_c^*}=f_{B_c}=517\,\rm{MeV}$ from the Richardson¡¯s potential \cite{Fulc};
  $f_{B_c}=433\,\rm{MeV}$ and  $f_{B_c^*}=503\,\rm{MeV}$ from the relativistic quark model with an special potential \cite{EFG};
   $f_{B_c}=(410\pm 40)\,\rm{MeV}$ from the relativized quark (Godfrey-Isgur) model \cite{Godfrey2004};
   $f_{B_c}=(420\pm 13)\,\rm{MeV}$ from the lattice non-relativistic QCD \cite{Jones-latt};
 $f_{B_c}=(395\pm15)\,\rm{MeV}$ from the QCD-motivated potential model \cite{Kiselev04};
  $f_{B_c^*}=f_{B_c}=315^{+26}_{-50} \,\rm{MeV}$ from the  shifted
$N$-expansion method \cite{Ikhdair2006};
  $f_{B_c}=377\,\rm{MeV}$ ($360\,\rm{MeV}$, $440\,\rm{MeV}$), $f_{B_c^*}=398\,\rm{MeV}$ ($387\,\rm{MeV}$, $440\,\rm{MeV}$) from the light-front quark model \cite{ChoiJ}(\cite{Hwang},\cite{RCVerma});
   $f_{B_c^*}=(453\pm20)\,\rm{MeV}$, $f_{B_c}=(438\pm 10)\,\rm{MeV}$   from the field correlator method \cite{BBS};
   $f_{B_c}=(322 \pm 42)\,\rm{MeV}$, $f_{B_c^*}=(418 \pm 24)\,\rm{MeV}$ from the Bethe-Salpeter equation \cite{WangGL}.

Although the values of the decay constants vary in large ranges, some theoretical calculations indicate that the decay constants have the relation $f_{B_c^*} \approx ({\rm or = }) f_{B_c}$ \cite{Colangelo1993,EQ,Fulc,Ikhdair2006,ChoiJ,Hwang,RCVerma,BBS}. In the early work \cite{Khlopov-2},   Gershtein  and
 Khlopov obtained a simple relation $f_{ij}\propto m_i+m_j$ for the decay constant $f_{ij}$ of the pseudoscalar meson having the constituent quarks $i$ and $j$,
 such simple relation does not work well enough  for both the light and heavy quarks.
In this article, we choose the values $f_{B_c^*}=0.384 \,\rm{GeV}$, $M_{B_c^*}=
6.337 \,\rm{GeV}$ from the recent analysis based on  the QCD sum rules \cite{Wang-BC}, $f_{B_c}=395 \rm{MeV}$ from the QCD-motivated potential model \cite{Kiselev04}, $M_{B_c}=6.277\,\rm{GeV}$ from the Particle Data Group \cite{PDG}.
The decay constants have the relation $f_{B_c^*} \approx f_{B_c}$,   the masses have the splitting $M_{B_c^*}-M_{B_c}=60\,\rm{MeV}$. The uncertainties of the electromagnetic form-factor $V(q^2)$ originate from
the decay constants can be estimated as $ \frac{\delta f_{B_c^*}}{f_{B_c^*}} + \frac{\delta f_{B_c}}{f_{B_c}}$. The calculations based on the nonrelativistic renormalization group indicate that $M_{B_c(1^-)}-M_{B_c(0^-)}=(50 \pm 17  {}^{+15}_{-12})\,\rm{MeV}$
 \cite{Penin2004}, the mass $M_{B_c^*}=6.337\,\rm{GeV}$ from the QCD sum rules is satisfactory. Accordingly, we take the threshold parameters and Borel  parameters
as  $s_1^0=s_2^0=(45\pm1)\,\rm{GeV}^2$,  $M_1^2=M_2^2=(5-7)\,\rm{GeV}^2$ from the QCD sum rules \cite{Wang-BC}.

The value of the gluon condensate $\langle \frac{\alpha_s
GG}{\pi}\rangle $ has been updated from time to time, and changes
greatly, we use the recently updated value $\langle \frac{\alpha_s GG}{\pi}\rangle=(0.022 \pm
0.004)\,\rm{GeV}^4 $ \cite{gg-conden}.
For the heavy quark masses, we take the $\overline{MS}$ masses $m_{c}(m_c^2)=(1.275\pm0.025)\,\rm{GeV}$ and  $m_{b}(m_b^2)=(4.18\pm 0.03)\,\rm{GeV}$
 from the Particle Data Group \cite{PDG}, and take into account
the energy-scale dependence of  the $\overline{MS}$ masses from the renormalization group equation,
\begin{eqnarray}
m_c(\mu^2)&=&m_c(m_c^2)\left[\frac{\alpha_{s}(\mu)}{\alpha_{s}(m_c)}\right]^{\frac{12}{25}} \, ,\nonumber\\
m_b(\mu^2)&=&m_b(m_b^2)\left[\frac{\alpha_{s}(\mu)}{\alpha_{s}(m_b)}\right]^{\frac{12}{23}} \, ,\nonumber\\
\alpha_s(\mu)&=&\frac{1}{b_0t}\left[1-\frac{b_1}{b_0^2}\frac{\log t}{t} +\frac{b_1^2(\log^2{t}-\log{t}-1)+b_0b_2}{b_0^4t^2}\right]\, ,
\end{eqnarray}
  where $t=\log \frac{\mu^2}{\Lambda^2}$, $b_0=\frac{33-2n_f}{12\pi}$, $b_1=\frac{153-19n_f}{24\pi^2}$, $b_2=\frac{2857-\frac{5033}{9}n_f+\frac{325}{27}n_f^2}{128\pi^3}$,  $\Lambda=213\,\rm{MeV}$, $296\,\rm{MeV}$  and  $339\,\rm{MeV}$ for the flavors  $n_f=5$, $4$ and $3$, respectively  \cite{PDG}. In this article, we take the typical energy scale $\mu=  2\,\rm{GeV}$ as in Ref.\cite{Wang-BC}.

\begin{figure}
 \centering
 \includegraphics[totalheight=4cm,width=5cm]{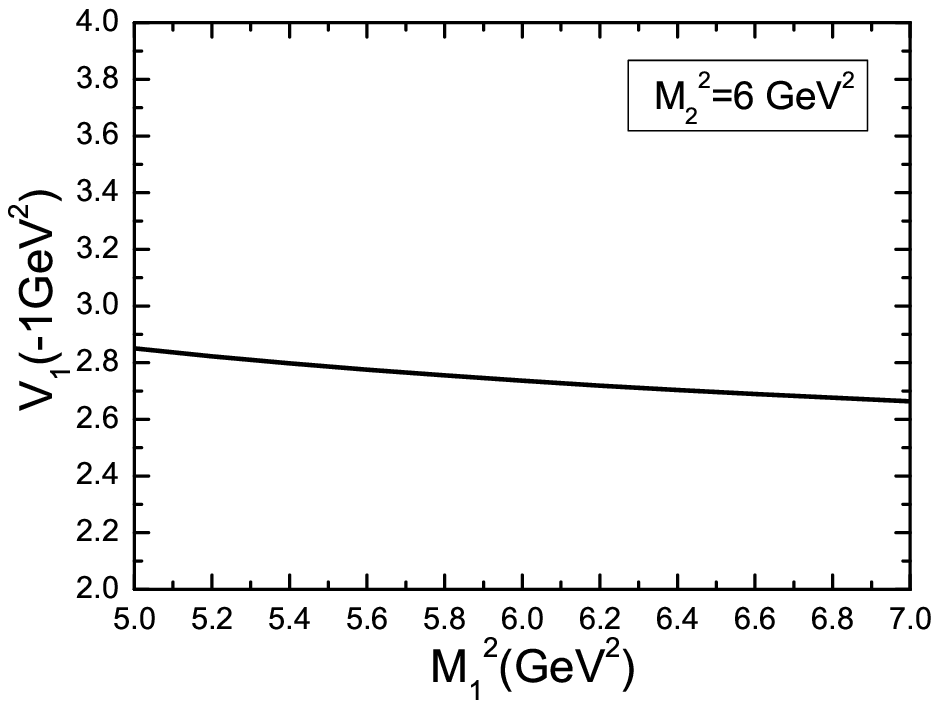}
 \includegraphics[totalheight=4cm,width=5cm]{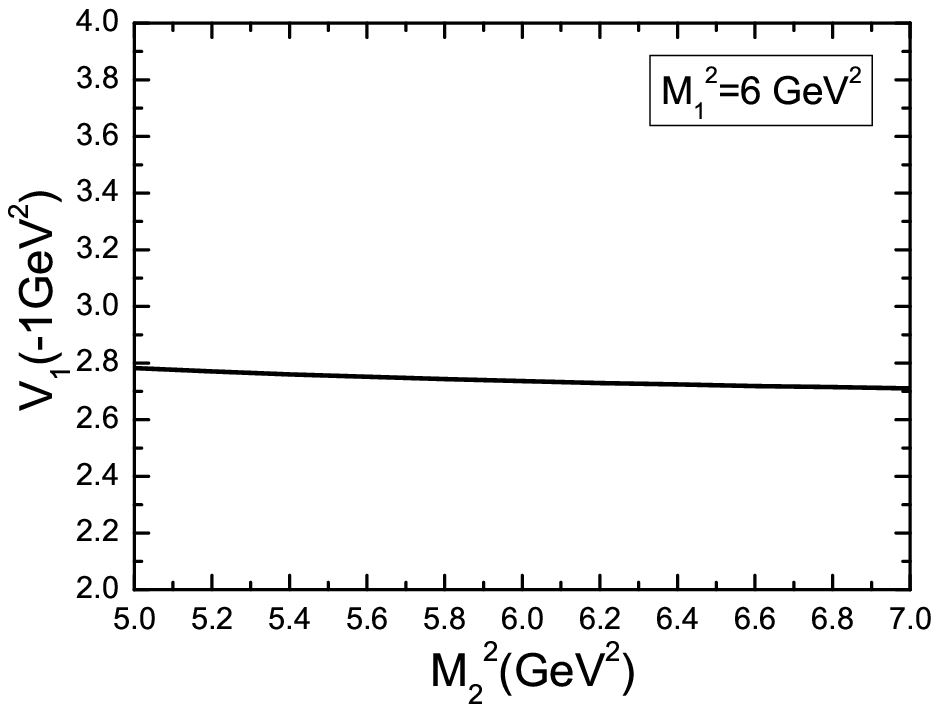}
 \includegraphics[totalheight=4cm,width=5cm]{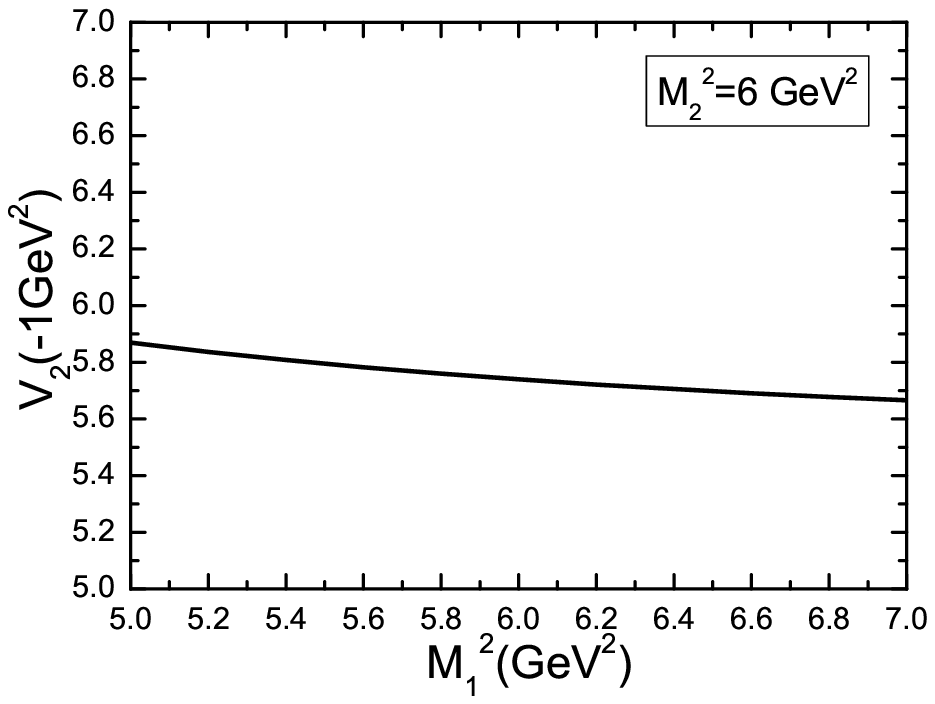}
 \includegraphics[totalheight=4cm,width=5cm]{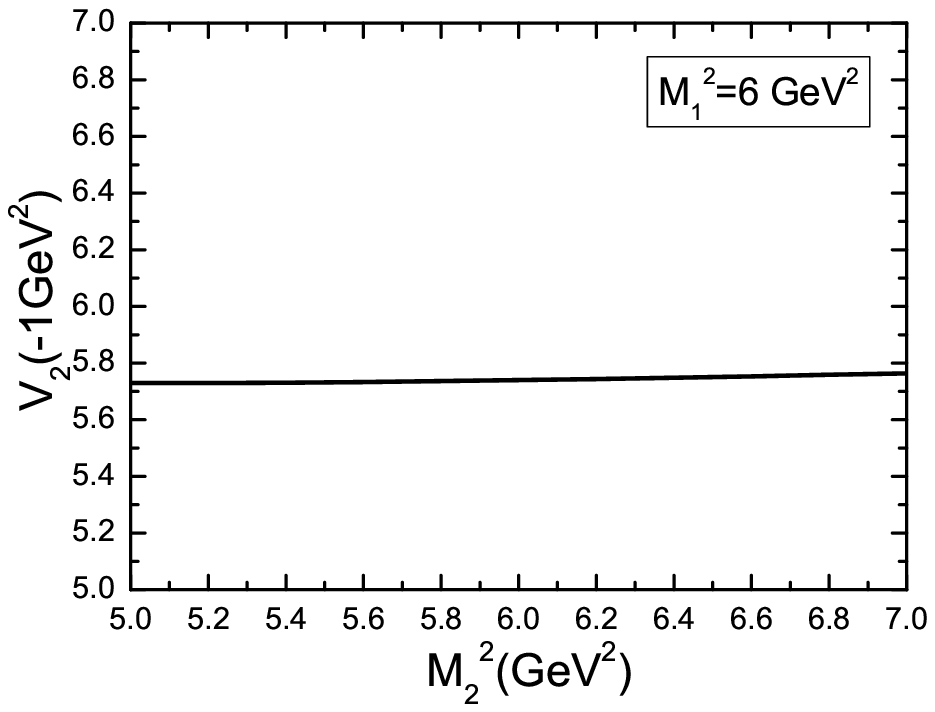}
        \caption{ The electromagnetic form-factors $V_1(q^2=-1\,\rm{GeV}^2)$ and $V_2(q^2=-1\,\rm{GeV}^2)$ with variations of the Borel parameters $M_1^2$ and $M_2^2$. }
\end{figure}

In Fig.2, we plot the electromagnetic form-factors at $q^2=-Q^2=-1\,\rm{GeV}^2$ with variations of the Borel parameters $M_1^2$ and $M_2^2$. From the figure, we can see that the  values are rather stable with variations of the  Borel parameters.  In calculations, we observe that $0.0001\leq \exp(-\frac{s_1^0}{M_1^2})\leq 0.00186$ and $0.0001\leq \exp(-\frac{s_2^0}{M_2^2})\leq 0.00186 $, the contributions from high resonances and continuum states are greatly suppressed, furthermore, the contributions from the gluon condensate are of minor importance, the operator product expansion is well convergent.

We take into account all the uncertainties  come from the input parameters, such as the heavy quark masses, threshold parameters, Borel parameters, $\dots$, obtain numerical values of the electromagnetic form-factors $V_1(Q^2)$, $V_2(Q^2)$ and $V(Q^2)$ from Eqs.(12-13), and  show them explicitly in Figs.3-4. We express the electromagnetic form-factors in the standard form $f(Q^2)=\overline{f}(Q^2)\pm \delta f(Q^2) $ numerically, where the $f(Q^2)$ denotes the electromagnetic form-factors  $V_1(Q^2)$, $V_2(Q^2)$, $V(Q^2)$, the $\overline{f}(Q^2)$ denotes the central values, and the $\delta f(Q^2)$ denotes the uncertainties, then fit the numerical values of the $V_1(Q^2)$, $V_2(Q^2)$ at $Q^2=(1-5.4)\,\rm{GeV}^2$ and $V(Q^2)$ at $Q^2=(1-4.2)\,\rm{GeV}^2$  into the following  analytical functions,
 \begin{eqnarray}
  V_1(Q^2)&=&\frac{A}{1+BQ^2}\, , \nonumber\\
  V_2(Q^2)&=&\frac{C}{1+DQ^2+EQ^4}\exp\left( -FQ^2\right)\, ,\\
    V(Q^2)&=&G\exp\left(-HQ^2 \right)+T\, ,
  \end{eqnarray}
  with the $\bf \rm MINUIT$, and determine the parameters,
    \begin{eqnarray}
    A&=&2.8905\pm 0.45717 \, ,\nonumber\\
    B&=&0.056340\pm 0.056316\,\rm{GeV}^{-2} \, ,\nonumber\\
    C&=&10.978\pm 10.369 \, ,\nonumber\\
    D&=&0.20611\pm 0.94270 \,\rm{GeV}^{-2}\, ,\nonumber\\
    E&=&0.017546\pm 0.55116\,\rm{GeV}^{-4} \, ,\nonumber\\
    F&=&0.44543\pm 1.4035 \,\rm{GeV}^{-2}\, ,\nonumber\\
    G&=&7.0807\pm 1.8756\,, \nonumber\\
    H&=&0.67821\pm  0.22875  \,\rm{GeV}^{-2}\, ,\nonumber\\
    T&=&-0.66869\pm 0.35949\, .
    \end{eqnarray}
From Figs.3-4, we can see that the  fitted functions can reproduce the central values of the form-factors at large ranges $Q^2=(1-10)\,\rm{GeV}^2$,
and the fitted functions $V_1(Q^2)$, $V_2(Q^2)$ and $V(Q^2)$ work well.

 We continue the $Q^2$ to the physical region $Q^2=0$ analytically to obtain  the physical electromagnetic form-factor $V(0)$,
\begin{eqnarray}
 V(0)&=&6.35517 \pm 6.91435 \, \, \,   {\rm from \, \, \, Eq.(18)}\, ,\nonumber\\
     &=&6.41201\pm1.90974 \, \, \,   {\rm from \, \, \, Eq.(19)}\, .
\end{eqnarray}
The curve of the fitted function $V_2(Q^2)$ is very steep, the value $V_2(0)=10.978\pm 10.369$  has too large uncertainty,
the resulting uncertainty of the $V(0)$ is also too large, we discard the value $ V(0)=6.35517 \pm 6.91435$. On the other hand, the value $V(0)=6.41201\pm1.90974$ from
Eq.(19) has much smaller uncertainty, i.e. less than $30\%$. We take the value $ V(0)=6.41201\pm1.90974$, and obtain
the radiative decay width,
\begin{eqnarray}
\Gamma(B_c^*\to B_c \gamma) &=&\frac{\alpha |V(0)|^2\left(M_{B_c^*}+M_{B_c}\right)\left(M_{B_c^*}-M_{B_c}\right)^3}{24M_{B_c^*}^3}    \nonumber\\
&=&133.9^{+91.6}_{-67.9}\,\rm{eV} \, \left(133.9 \pm 79.7\,\rm{eV} \right) \, ,
\end{eqnarray}
where the fine constant $\alpha=\frac{1}{137}$, the asymmetric uncertainty comes  from the formula    $V^2(0)-\overline{V}^2(0)$, while the symmetric uncertainty in the bracket  comes  from the approximation $V^2(0)-\overline{V}^2(0)\approx \pm 2\overline{V}(0)\delta V(0)$  with $V(0)=\overline{V}(0)\pm \delta V(0)$. From Eq.(22) we can see that the decay width is sensitive to the mass splitting $M_{B_c^*}-M_{B_c}$ as $\Gamma(B_c^*\to B_c \gamma)\propto (M_{B_c^*}-M_{B_c})^3$.
The present prediction $\Gamma(B_c^* \to B_c \gamma)=133.9^{+91.6}_{-67.9}\,\rm{eV}\, \left(133.9 \pm 79.7\,\rm{eV} \right)$ is compatible with previous values $60\,\rm{eV}$ from the nonrelativistic potential \cite{GKLT}, $134.5\,\rm{eV}$ from non-relativistic  potential  model \cite{EQ}, $59\,\rm{eV}$ from Richardson¡¯s potential \cite{Fulc},
 $33\,\rm{eV}$ from the relativistic quark model with an special potential \cite{EFG},  $80\,\rm{eV}$ from the relativized quark  (Godfrey-Isgur) model \cite{Godfrey2004}.
In Ref.\cite{Wang1209}, we have used a larger decay constant $f_{B_c^*}=0.79\,\rm{GeV}$ rather than $0.384\,\rm{GeV}$, the smaller decay constant $f_{B_c^*}=0.384\,\rm{GeV}$ leads  to the semileptonic decay widths $\Gamma(B_c^*\to \eta_c \ell \bar{\nu}_\ell)\sim 10^{-5}\,\rm{eV}$.
  The  branching  fractions of the semileptonic decays $B_c^*\to \eta_c \ell \bar{\nu}_\ell$ are of the order $10^{-7}\sim 10^{-6}$, which supports that the dominant decay model is $B_c^* \to B_c \gamma$. We can search for the $B_c^*$ mesons using  the decay cascades  $B_c^{*\pm}\to B_c^{\pm} \gamma\to J/\psi \ell^{\pm}\bar{\nu}_{\ell} \gamma\to \mu^+ \mu^- \ell^{\pm}\bar{\nu}_{\ell} \gamma$.

\begin{figure}
 \centering
 \includegraphics[totalheight=6cm,width=7cm]{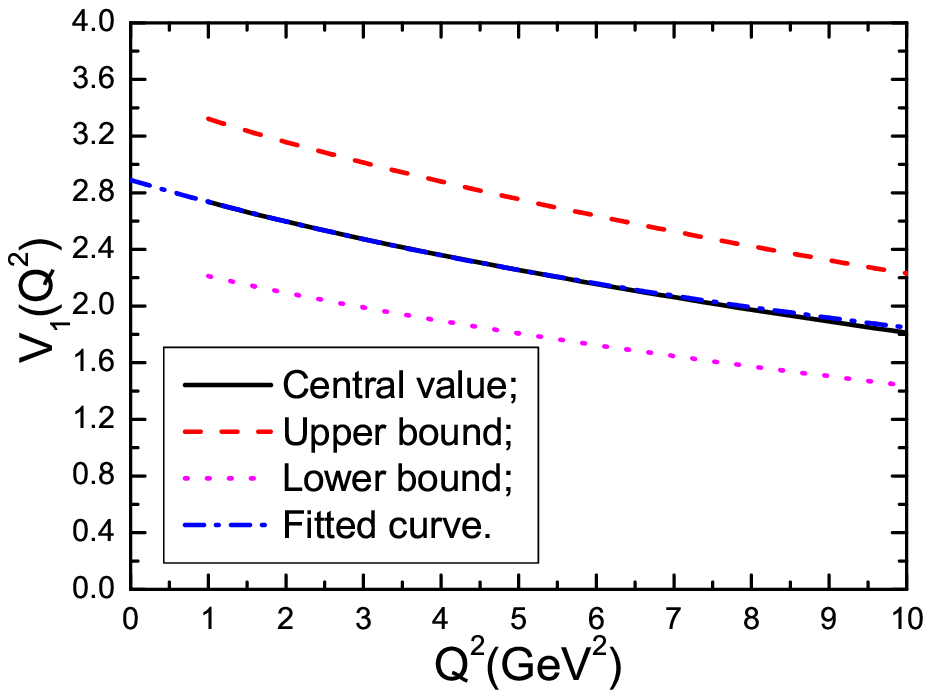}
\includegraphics[totalheight=6cm,width=7cm]{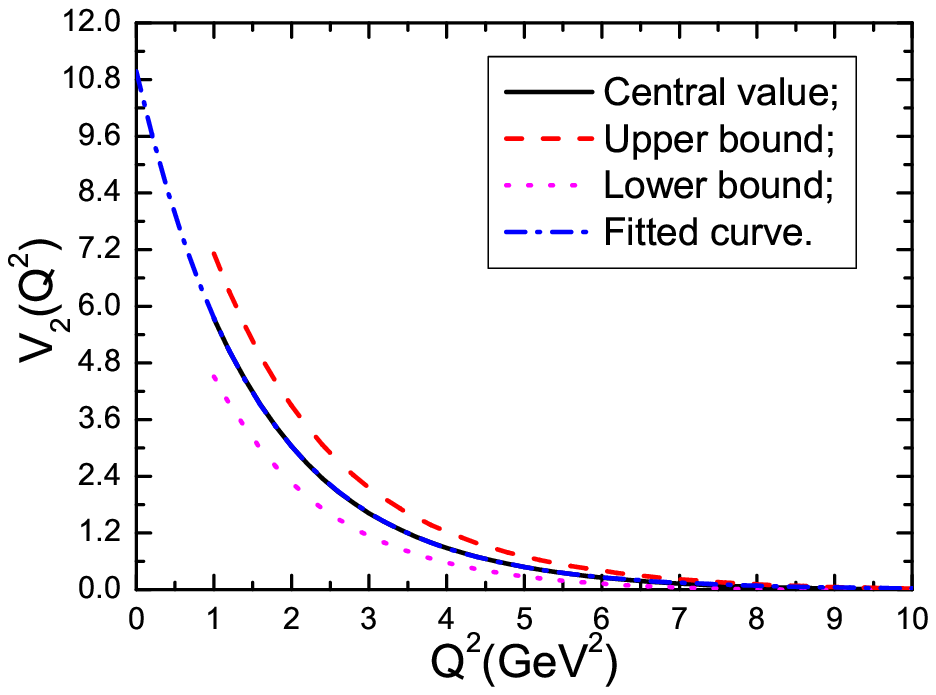}
        \caption{ The electromagnetic form-factors $V_1(Q^2)$ and $V_2(Q^2)$, where the "Fitted curve" denotes the central values of the fitted functions. }
\end{figure}

\begin{figure}
 \centering
 \includegraphics[totalheight=7cm,width=9cm]{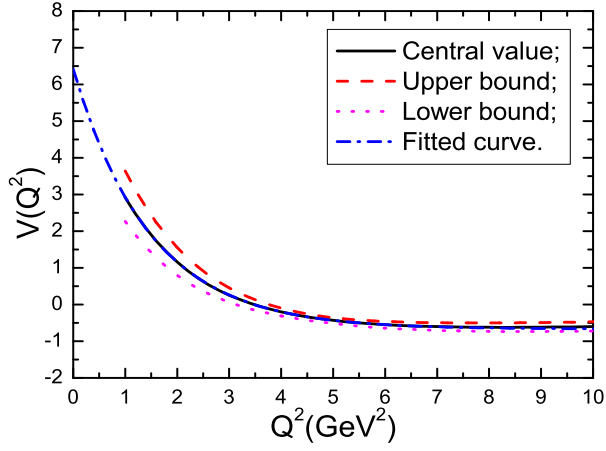}
        \caption{ The electromagnetic form-factor $V(Q^2)$, where the "Fitted curve" denotes the central value of the fitted function. }
\end{figure}

\section{Conclusion}
In this article, we calculate the $B_c^* \to B_c$ electromagnetic form-factor with the three-point QCD sum rules, and obtain the numerical values for the form-factor at momentum transfer $Q^2=-q^2=(1.0-5.4)\,\rm{GeV}^2$, then fit the form-factors to analytical functions  to obtain the physical value, and study  the radiative decays $B_c^{*\pm} \to B_c^{\pm} \gamma$.     We expect to study the radiative transitions    using the decay cascades  $B_c^{*\pm}\to B_c^{\pm} \gamma\to J/\psi \ell^{\pm}\bar{\nu}_{\ell} \gamma\to \mu^+ \mu^- \ell^{\pm}\bar{\nu}_{\ell} \gamma$   in the future at the LHCb.

\section*{Acknowledgements}
This  work is supported by National Natural Science Foundation,
Grant Number 11075053,  and the Fundamental Research Funds for the
Central Universities.

\section*{Appendix}
The explicit expressions of the $I_0^{ijn}$, $I_{10}^{ijn}$, $I_{01}^{ijn}$, $J_0^{ijn}$, $J_{10}^{ijn}$, $J_{01}^{ijn}$,
\begin{eqnarray}
iI_0^{ijn}&=&B_{-p_1^2\rightarrow M_1^2}B_{-p_2^2\rightarrow M_2^2} \overline{I}_{ijn} \nonumber\\
&=&B_{P_1^2\rightarrow M_1^2}B_{P_2^2\rightarrow M_2^2}\frac{(-1)^{i+j+n}i}{\Gamma(i)\Gamma(j)\Gamma(n)}\int d^4K \int_0^\infty d\alpha d\beta d\gamma \alpha^{i-1}\beta^{j-1}\gamma^{n-1} \nonumber\\
&&\exp\left\{ -\alpha(K+P_1)^2-\beta(K+P_2)^2-\gamma K^2-\alpha m_b^2-\beta m_b^2-\gamma m_c^2\right\} \nonumber\\
&=&\frac{(-1)^{i+j+n}i\pi^2}{\Gamma(i)\Gamma(j)\Gamma(n)(M_1^2)^i(M_2^2)^j (M^2)^{n-2}}\int_0^1 d\lambda \frac{\lambda^{1-i-j}}{(1-\lambda)^{n-1}} \nonumber\\
&&\exp\left\{ -\frac{(1-\lambda)Q^2}{\lambda(M_1^2+M_2^2)}-\frac{m_b^2}{\lambda M^2}-\frac{m_c^2}{(1-\lambda)M^2}\right\}\nonumber\\
&=&\frac{(-1)^{i+j+n}i\pi^2}{\Gamma(i)\Gamma(j)\Gamma(n)(M_1^2)^i(M_2^2)^j (M^2)^{n-2}}\int_0^\infty d\tau (\tau+1)^{i+j+n-4}\tau^{1-i-j} \nonumber\\
&&\exp\left\{ -\frac{ 1}{\tau}\left(\frac{Q^2}{M_1^2+M_2^2}+\frac{m_b^2}{M^2}\right)-\frac{m_b^2+m_c^2}{ M^2}-\tau\frac{m_c^2}{M^2}\right\} \, ,
\end{eqnarray}

\begin{eqnarray}
iI^\mu_{ijn}&=&B_{-p_1^2\rightarrow M_1^2}B_{-p_2^2\rightarrow M_2^2} \overline{I}^\mu_{ijn} \nonumber\\
&=&\frac{(-1)^{i+j+n+1}i\pi^2}{\Gamma(i)\Gamma(j)\Gamma(n)(M_1^2)^{i+1}(M_2^2)^j (M^2)^{n-3}}\int_0^\infty d\tau (\tau+1)^{i+j+n-3}\tau^{1-i-j} \nonumber\\
&&\exp\left\{ -\frac{ 1}{\tau}\left(\frac{Q^2}{M_1^2+M_2^2}+\frac{m_b^2}{M^2}\right)-\frac{m_b^2+m_c^2}{ M^2}-\tau\frac{m_c^2}{M^2}\right\}p_{1}^\mu \nonumber\\
&&+\frac{(-1)^{i+j+n+1}i\pi^2}{\Gamma(i)\Gamma(j)\Gamma(n)(M_1^2)^i(M_2^2)^{j+1} (M^2)^{n-3}}\int_0^\infty d\tau (\tau+1)^{i+j+n-3}\tau^{1-i-j} \nonumber\\
&&\exp\left\{ -\frac{ 1}{\tau}\left(\frac{Q^2}{M_1^2+M_2^2}+\frac{m_b^2}{M^2}\right)-\frac{m_b^2+m_c^2}{ M^2}-\tau\frac{m_c^2}{M^2}\right\}p_{2}^\mu \nonumber\\
&=&iI_{10}^{ijn}p_{1}^\mu+iI_{01}^{ijn}p_{2}^\mu\, ,
\end{eqnarray}

\begin{eqnarray}
J_0^{ijn}&=&I_0^{ijn}|_{m_b\leftrightarrow m_c} \, ,\nonumber\\
J_{10}^{ijn}&=&I_{10}^{ijn}|_{m_b\leftrightarrow m_c} \, ,\nonumber\\
J_{01}^{ijn}&=&I_{01}^{ijn}|_{m_b\leftrightarrow m_c} \, , \nonumber\\
M^2&=&\frac{M_1^2M_2^2}{M_1^2+M_2^2} \, ,
\end{eqnarray}
where we have used the Borel transform $B_{P^2\rightarrow M^2}\exp(-\alpha P^2)=\delta(1-\alpha M^2)$. Those analytical  expressions
are slightly  different from that obtained in Ref.\cite{Kiselev2000}, they are both correct.

\end{document}